\RequirePackage{fix-cm}
\documentclass[smallextended]{svjour3}       
\smartqed  
\usepackage{graphicx}
\usepackage{amssymb}
\usepackage{amsmath}
\usepackage{hhline}

\def\bra#1{\mathinner{\langle{#1}|}}
\def\ket#1{\mathinner{|{#1}\rangle}}
\def\braket#1{\mathinner{\langle{#1}\rangle}}

\begin{document}

\title{The Heun differential equation and the Gauss differential equation related to quantum walks
}
\subtitle{}

\titlerunning{The relations between quantum walks and differential equations}        

\author{Norio Konno \and Takuya Machida \and Tohru Wakasa}




\institute{
\begin{acknowledgements}
N. K. was partially supported by the Grant-in-Aid for Scientific Research (C) of Japan Society for the Promotion of Science (Grant No. 21540118).
T. M. and T. W. are grateful to the Meiji University Global COE Program ``Formation and Development of Mathematical Sciences Based on Modeling and Analysis'' for the support.
\end{acknowledgements}
}

\date{}

\maketitle
\vspace{5mm}

\begin{abstract}
The limit theorems of discrete- and continuous-time quantum walks on the line have been intensively studied.
We show a relation among limit distributions of quantum walks, Heun differential equations and Gauss differential equations.
Indeed, we derive the second-order Fucksian differential equations which limit density functions of quantum walks satisfy.
Moreover, using both differential equations, we discuss a relationship between discrete- and continuous-time quantum walks.
Taking suitable limit, we can transform a Heun equation obtained from the limit density function of the discrete-time quantum walk to a Gauss equation given by that of the continuous-time quantum walk.
\end{abstract}

\section{Introduction}
\label{intro}
The discrete-time quantum walk (QW), which is a quantum counterpart of the classical random walk, has been extensively investigated since Ambainis et al.~\cite{ambainis2001one} studied a detail of the walk.
The continuous-time QW was proposed by Farhi and Gutmann~\cite{farhi1998quantum}, and has been analyzed on not only regular graphs but also complex networks~(e.g.~\cite{ide2010continuous,mulken2007quantum,xu2008continuous}).
A further development in the theory of the QW during recent 10 years showed us several novel properties on both QWs. 
The behavior of the QW is quite different from that of the random walk and is expected to be connected with various phenomena in quantum mechanics. 
A relation between the QW and quantum computer has been also discussed~(e.g.~\cite{ambainis2004quantum,nielsen2000quantum,shenvi2003quantum}).
The quantum search algorithm designed by the QW is one of the important applications.
The Grover search algorithm can be considered as a discrete-time QW on the complete graph and it produces speed-up at the square-root rate for the corresponding classical search.
There are some reviews of the QW~\cite{kempe2003quantum,kendon2007decoherence,konno2008quantum,venegas2008quantum}.

One of the purposes in the study of the QW is to derive the limit distribution and the asymptotic behavior as time tends to infinity.
The limit distribution has been obtained for various kinds of QWs.
In the present paper, we concentrate on the QW in the one-dimensional space.
Observing previous results, one can mention that each limit distribution usually has a compact support and admits a singularity at the boundary of the support.
This is an interesting property of the limit distribution, because the QW does not possess any singularity in space.
To the fact, we now address a question:
How to understand the singularity of the QW?
Motivated by this question, we investigate the limit distribution of the QW from a viewpoint of the differential equation.

We begin with finding a differential equation for the limit distribution.
For example, consider discrete- and continuous-time symmetric simple random walks on the line.
The central limit theorem for the walk, that is, $\lim_{t\to\infty}\mathbb{P}(Y_t/\sqrt{t}\leq x)=\int_{-\infty}^x e^{-y^2/2}/\sqrt{2\pi}\,dy$, is well-known, where $Y_t$ denotes the walker's position at time $t$ and $\mathbb{P}(Y_t=x)$ is the probability that the walker is at position $x$ at time $t$.
For the limit density function $f(x)=e^{-x^2/2}/\sqrt{2\pi}$, an equation $d^2f(x)/dx^2+xdf(x)/dx+f(x)=0$ is derived.
This equation comes from the theory of the diffusion equation $\partial u(x,t)/\partial t=\frac{1}{2}\partial^2 u(x,t)/\partial x^2$, and in particular, it is also obtained by the {\it self-similarity} of the fundamental solution $u(x,t)=(1/\sqrt t) f(x/\sqrt t)$.

Let us consider both discrete-time and continuous-time QWs on the line.
The classical random walk corresponds to the diffusion process, while for the QW the corresponding process is not known.
Thus, to find out the process would remain as one of important problems.
Taking an account of the singularity in the limit distribution, we treat a class of Fucksian linear differential equations of the second order.
For discrete-time (resp. continuous-time) QWs, we are led to a Heun's differential equation (HE) (resp. a hypergeometric differential equation by Gauss (the Gauss equation, GE)).
The GE is one of typical Fucksian equations and represents the Fucksian equations with exactly three regular singular points.
On the other hand, the HE was proposed in 1888 by Heun~\cite{heun1888theorie} and 
admits four regular singular points. 

Moreover, a concept of confluence between the HE and the GE helps us to understand a relationship between discrete- and continuous-time QWs clearly. 
Through these results, authors believe that the HE and the GE play an important role in understanding the QWs.

To this end, we will mention a significant remark on the QW and the $BC_1$ Inozemtsev model, which is an integrable quantum system.
These two models are to be connected through HEs.
Therefore, we would like to expect a new mathematical theory on the relation between QWs, HEs, and quantum mechanics.

The present paper is organized as follows.
In Sect.~\ref{dt}, we explain both the one-dimensional discrete-time QW and the HE.
After introducing the limit distribution obtained by Konno~\cite{konno2002quantum,konno2005new}, we discuss a relation between the discrete-time QW and the HE.
In Sect.~\ref{ct}, we concentrate on the continuous-time QW, and show that the GE relates with the limit density function of the walk. 
Section~\ref{dtct} is devoted to a connection between the HE and the GE in order to understand a relationship between discrete- and continuous-time QWs given by Strauch~\cite{strauch2006connecting}.
In the final section, we summarize the relations obtained here and propose a future problem on our results.
Furthermore, we transform the HE to the $BC_1$ Inozemtsev system in \ref{app1} and show relations among measure of the QW, the HE and the GE in \ref{app2}.


\section{A relation between the discrete-time QW and the HE}
\label{dt}

In this section, we discuss a relation between the discrete-time QW in one dimension and the HE. 
At first we define the QW on the line.
Let $\ket{x}$ ($x\in\mathbb{Z}=\left\{0,\pm 1,\pm2,\ldots\right\}$) be infinite components vectors which denote the position of the walker.
Here,  $x$-th component of $\ket{x}$ is 1 and the other is 0.
Let $\ket{\psi_{t}(x)} \in \mathbb{C}^2$ be the amplitude of the walker at position $x$ at time $t \in\left\{0,1,2,\ldots\right\}$, where $\mathbb{C}$ is the set of complex numbers.
The amplitude of the walk at time $t$ is expressed by
\begin{equation}
 \ket{\Psi_t}=\sum_{x\in\mathbb{Z}}\ket{x}\otimes\ket{\psi_{t}(x)}.
\end{equation}
The time evolution of the walk can be defined by the following unitary matrix:
\begin{equation}
  U=\left[\begin{array}{cc}
    a & b \\ c & d
	 \end{array}\right],
\end{equation}
where $a,b,c,d\,\in\mathbb{C}$.
Moreover, we introduce two matrices:
\begin{equation}
 P=\left[\begin{array}{cc}
    a & b\\ 0&0
	 \end{array}\right],\,
 Q=\left[\begin{array}{cc}
    0&0\\ c & d
	 \end{array}\right].
\end{equation}
Note that $P+Q=U$.
Then the evolution is determined by
\begin{equation}
 \ket{\psi_{t+1}(x)}=P\ket{\psi_t(x+1)}+Q\ket{\psi_t(x-1)}.\label{eq:te}
\end{equation}
\clearpage

\noindent The probability that the discrete-time quantum walker $X^{(d)}_t$ is at position $x$ at time $t$, $\mathbb{P}(X^{(d)}_t=x)$, is defined by
\begin{equation}
 \mathbb{P}(X^{(d)}_t=x)=\braket{\psi_t(x)|\psi_t(x)},\label{eq:prob}
\end{equation}
where $\bra{\psi_t(x)}$ denotes the conjugate transposed vector of $\ket{\psi_t(x)}$.
In the present paper, we take the initial state as
\begin{equation}
 \ket{\psi_0(x)}=\left\{\begin{array}{ll}
		 \!{}^T[\,1/\sqrt{2}, \,i/\sqrt{2}\,]& (x=0),\\
			\!{}^T[\,0,\,0\,]& (x\neq 0),
		       \end{array}\right.\label{eq_ini}
\end{equation}
where $T$ is the transposed operator.
Equation~(\ref{eq_ini}) is well-known as the initial state that gives the symmetric probability distribution about the origin~\cite{konno2008quantum}.

\begin{figure}[h]
 \begin{center}
 \begin{minipage}{60mm}
  \begin{center}
   \includegraphics[scale=0.35]{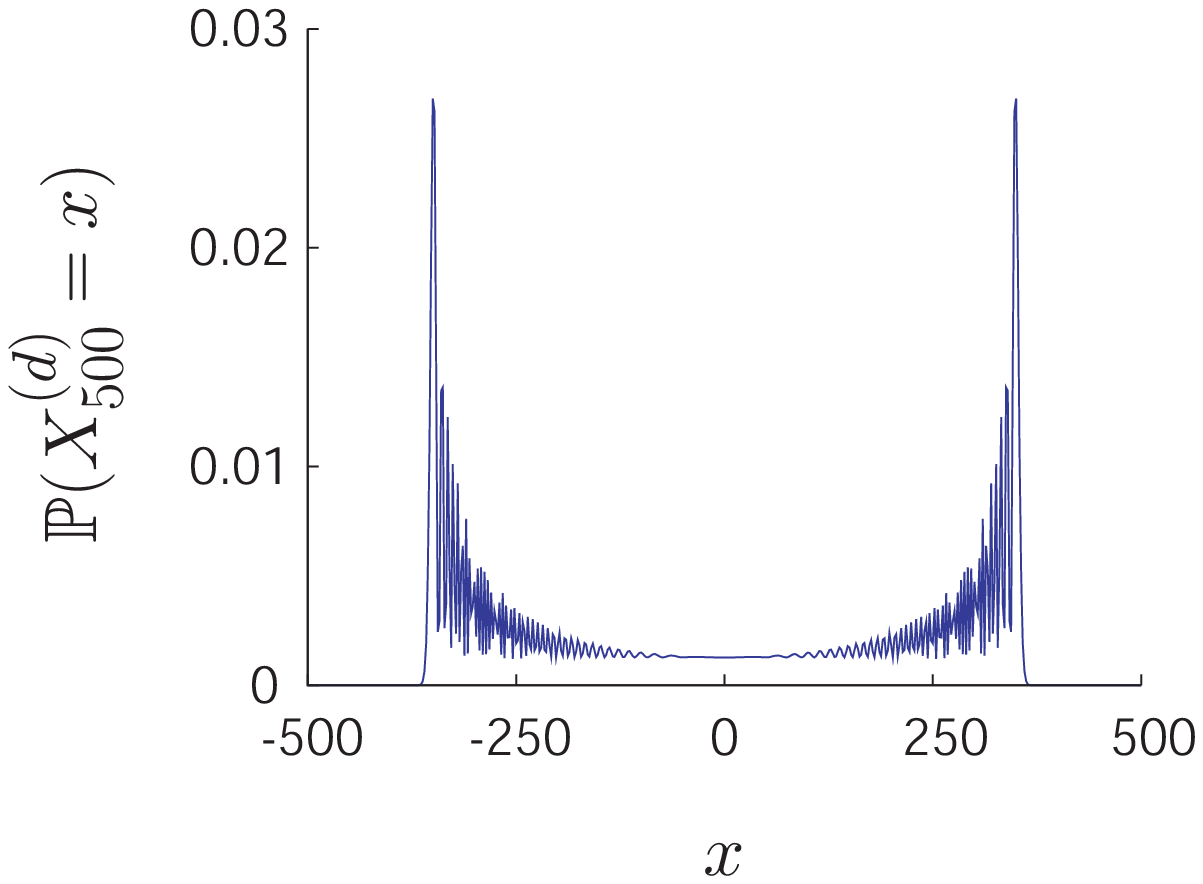}\\
   {(a)}
  \end{center}
 \end{minipage}
 \begin{minipage}{50mm}
  \begin{center}
   \includegraphics[scale=0.35]{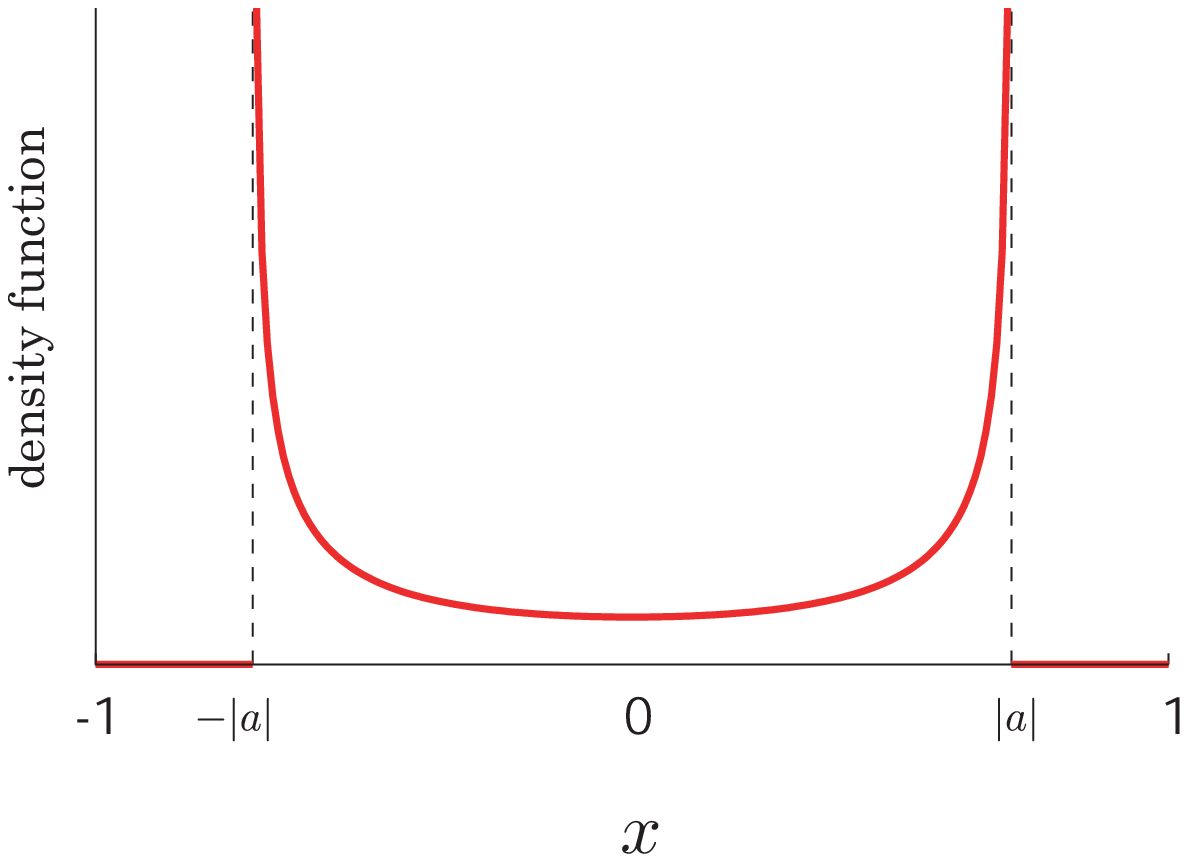}\\
   {(b)}
  \end{center}
 \end{minipage}
 \vspace{5mm}
 \caption{Comparison between the probability distribution and the limit density function of the discrete-time QW with $a=b=c=-d=1/\sqrt{2}$. Figure (a) is the probability distribution at time $t=500$. Figure (b) is the limit density function of the walk.}
  \label{fig:dtqw_distribution}
 \end{center}
\end{figure}

For the walk, some limit theorems and asymptotic behaviors have been obtained.
In particular, we focus on the density function of probability distribution as $t\to\infty$.  
The limit distribution of the QW was given by Konno~\cite{konno2002quantum,konno2005new}.
That is, for $abcd\neq 0$, we have
\begin{equation}
 \lim_{t\to\infty}\mathbb{P}\left(\frac{X^{(d)}_t}{t}\leq x\right)=\int_{-\infty}^x f^{(d)}(y)I_{(-|a|,|a|)}(y)\,dy,
\end{equation}
where
\begin{equation}
 f^{(d)}(x)=\frac{\sqrt{1-|a|^2}}{\pi(1-x^2)\sqrt{|a|^2-x^2}},
\end{equation}
and $I_A(x)=1$ if $x\in A$, $I_A(x)=0$ if $x\notin A$.
A lot of limit distributions of discrete-time QWs are often described by using this function $f^{(d)}(x)$ (e.g.~\cite{chisaki2009limit,inui2005one,konno2010limit,machida2010limit1,machida2010limit2,segawa2008limit}).
Figure~\ref{fig:dtqw_distribution} depicts the comparison between the probability distribution at time 500 and the limit density function for the walk with $a=b=c=-d=1/\sqrt{2}$, which is called the Hadamard walk.

From now, we discuss a relation between the limit density function $f^{(d)}(x)$ and the HE.
The canonical form of the HE is given by
\begin{equation}
 \frac{d^2 u}{dz^2}+\left(\frac{\gamma}{z}+\frac{\delta}{z-1}+\frac{\epsilon}{z-\theta}\right)\frac{du}{dz}+\frac{\alpha\beta z-q}{z(z-1)(z-\theta)}u=0,
\end{equation}
where $\theta \in\mathbb{C}$ is one of the singular points and $\alpha, \beta, \gamma, \delta, \epsilon, q \in\mathbb{C}$.
The five parameters $\alpha, \beta, \gamma, \delta, \epsilon$ are linked by the relation $\alpha+\beta+1=\gamma+\delta+\epsilon$ and the parameter $q$ is called the accessory parameter.
The HE has four regular singularities at $z=0,1,\theta,\infty$.
Particularly the HE with $\gamma=\delta=\epsilon=1/2$ is called Lam\'{e}'s equation and it have been analyzed in detail.
In some special cases, the HE becomes the GE (e.g. $\theta=1, q=\alpha\beta$ case).

We find that $f^{(d)}(x)$ satisfies the following differential equation:
\begin{align}
 (1-x^2)(|a|^2-x^2)\frac{d^2}{dx^2}f^{(d)}(x)-x(4|a|^2+3-7x^2)\frac{d}{dx}f^{(d)}(x)\nonumber\\
 +(9x^2-2|a|^2-1)f^{(d)}(x)=0.
\end{align}
By a change of an independent variable $x^2=z$, we obtain one of our main results:
\vspace{-1mm}
\begin{theorem}\label{th:dtqw}
\begin{align}
 \frac{d^2}{dz^2}u^{(d)}(z)+\left(\frac{\frac{1}{2}}{z}+\frac{2}{z-1}+\frac{\frac{3}{2}}{z-|a|^2}\right)\frac{d}{dz}u^{(d)}(z)\nonumber\\
 +\frac{\frac{9}{4}z-\frac{2|a|^2+1}{4}}{z(z-1)(z-|a|^2)}u^{(d)}(z)=0,\label{eq:dtqw_hde}
\end{align}
where
\begin{equation}
 u^{(d)}(z)=\frac{\sqrt{1-|a|^2}}{\pi(1-z)\sqrt{|a|^2-z}}.\label{eq:dtqw_hde_sol}
\end{equation}
\end{theorem}
We should remark that Eq.~(\ref{eq:dtqw_hde}) is equivalent to the HE with
\begin{equation}
 \alpha=\beta=\frac{3}{2},\, \gamma=\frac{1}{2},\, \delta=2,\, \epsilon=\frac{3}{2},\, q=\frac{2|a|^2+1}{4},\, \theta=|a|^2.
\end{equation}

\section{A relation between the continuous-time QW and the GE}
\label{ct}

In this section we will show a relationship between the continuous-time QW on $\mathbb{Z}$ and the GE:
\begin{equation}
 z(z-1)\frac{d^2 u}{dz^2}+\left\{(\alpha+\beta+1)z-\gamma\right\}\frac{du}{dz}+\alpha\beta u=0\label{eq:def_gde}
\end{equation}
where $\alpha, \beta, \gamma \in\mathbb{C}$ are parameters.
Equation~(\ref{eq:def_gde}) is a second-order Fucksian equation with three regular singularities at $z=0,1,\infty$.

\begin{figure}[h]
 \begin{center}
 \begin{minipage}{60mm}
  \begin{center}
   \includegraphics[scale=0.35]{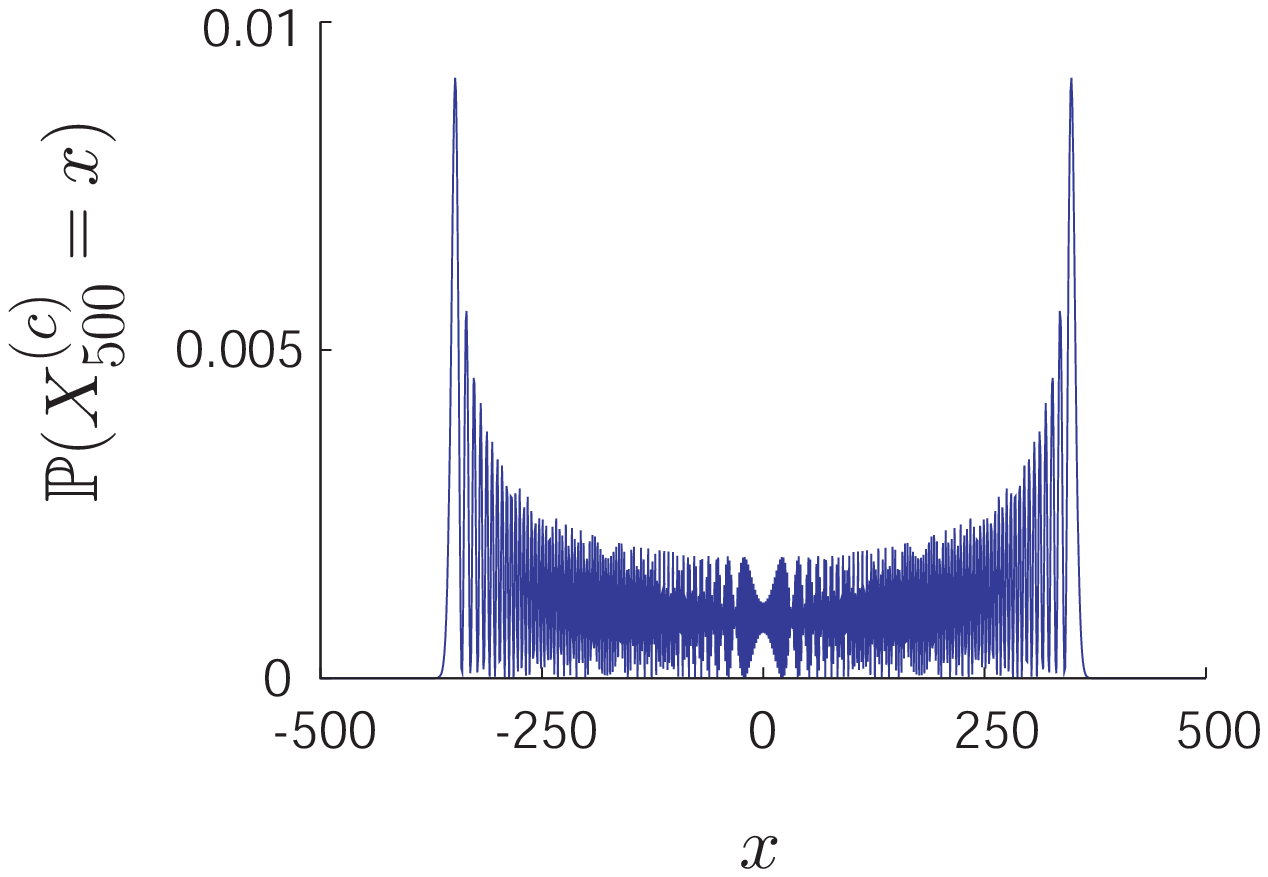}\\
   {(a)}
  \end{center}
 \end{minipage}
 \begin{minipage}{50mm}
  \begin{center}
   \includegraphics[scale=0.35]{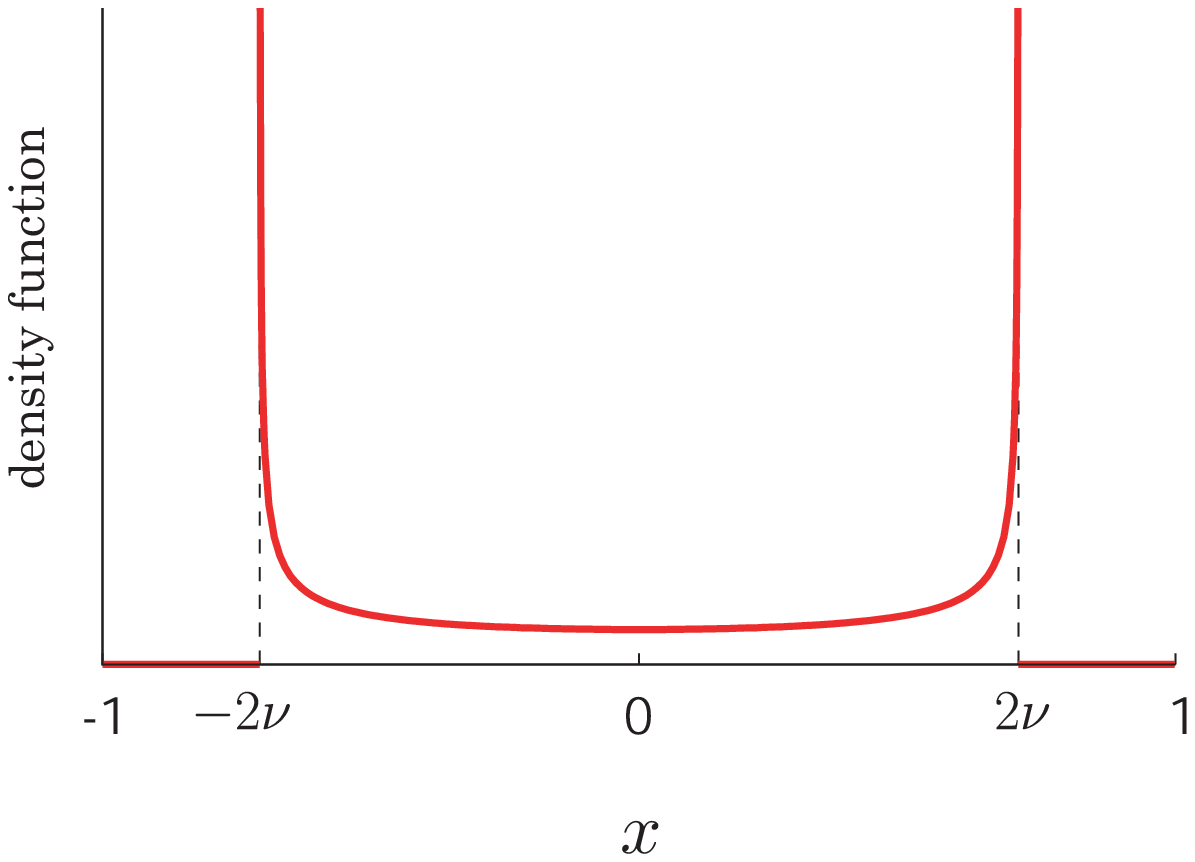}\\
   {(b)}
  \end{center}
 \end{minipage}
 \vspace{5mm}
 \caption{Comparison between the probability distribution and the limit density function of the continuous-time QW with $\nu=1/2\sqrt{2}$. Figure (a) is the probability distribution at time $t=500$. Figure (b) is the limit density function of the walk.}
  \label{fig:distribution}
 \end{center}
\end{figure}

We give a definition of the continuous-time QW on the line.
At first, we consider the amplitude $\psi_t(x)\in\mathbb{C}$ at position $x\in\mathbb{Z}$ at time $t\,(>0)$ instead of $\ket{\psi_t(x)}\in\mathbb{C}^2$ for the discrete-time QW.
The evolution of the amplitude is defined by
\begin{equation}
 i\frac{d\psi_t(x)}{dt}=-\nu\left\{\psi_t(x-1)-2\psi_t(x)+\psi_t(x+1)\right\},\label{eq:ctqw_te}
\end{equation}
where $\nu >0$.
The probability that the walker is at position $x$ at time $t$ is denoted by $\mathbb{P}(X^{(c)}_t=x)=|\psi_t(x)|^2$, where $X^{(c)}_t$ is the continuous-time quantum walker's position at time $t$.
We take an initial state as $\psi_0(0)=1$ and $\psi_0(x)=0\,(x\neq 0)$.
Konno \cite{konno2005limit} and Gottlieb \cite{gottlieb2005convergence} got the limit theorem for the continuous-time QW as follows:
\begin{equation}
 \lim_{t\rightarrow\infty}\mathbb{P}\left(\frac{X^{(c)}_t}{t}\leq x\right)=\int_{-\infty}^x f^{(c)}(y) \,I_{(-2\nu,2\nu)}(y)\,dy,
\end{equation}
where
\begin{equation}
 f^{(c)}(x)=\frac{1}{\pi\sqrt{(2\nu)^2-x^2}}.
\end{equation}
In Fig.~\ref{fig:distribution}, we show the comparison between the probability distribution at time 500 and the limit density function for the walk with $\nu=1/2\sqrt{2}$.

\noindent We see that the function $f^{(c)}(x)$ satisfies the following differential equation:
\begin{equation}
 (4\nu^2-x^2)\frac{d^2}{dx^2}f^{(c)}(x)-3x\frac{d}{dx}f^{(c)}(x)-f^{(c)}(x)=0.
\end{equation}
By the change of an independent variable $x^2/4\nu^2=z$, we have a continuous-time counterpart of Theorem~\ref{th:dtqw}:
\begin{theorem}
\begin{equation}
 z(z-1)\frac{d^2}{dz^2}u^{(c)}(z)+\left(2z-\frac{1}{2}\right)\frac{d}{dz}u^{(c)}(z)+\frac{1}{4}u^{(c)}(z)=0, \label{eq:ctqw_gde}
\end{equation}
where
\begin{equation}
 u^{(c)}(z)=\frac{1}{2\nu\pi\sqrt{1-z}}.\label{eq:ctqw_hde_sol}
\end{equation}
\end{theorem}
Note that Eq. (\ref{eq:ctqw_gde}) is the GE with $\alpha=\beta=\gamma=1/2$.

\section{A relation between discrete-time and continuous-time QWs in the theory of the differential equation}
\label{dtct}

In this section, we discuss a relation between discrete- and continuous-time QWs on the second-order differential equations which were obtained in previous sections.
Strauch~\cite{strauch2006connecting} found a relation between both walks by transforming Eq.~(\ref{eq:te}) to Eq.~(\ref{eq:ctqw_te}) in a suitable limit.
D'Alessandro~\cite{d2009connection} focused on both QWs on general graphs and derived the dynamics of continuous-time walks as a limit of discrete-time dynamics.

Considering a confluent type for the HE which the function $u^{(d)}(z)$ satisfies, we can also obtain the GE which the function $u^{(c)}(z)$ satisfies.  
For a scaling parameter $\tau$, substituting $z=t/\tau$ in Eq. (\ref{eq:dtqw_hde}), we have
\begin{align}
 4t(|a|^2\tau-t)\left(1-\frac{t}{\tau}\right)\frac{d^2}{dt^2}v(t)+2\left\{\frac{8t^2}{\tau}-(5|a|^2+4)t+|a|^2\tau\right\}\frac{d}{dt}v(t)\nonumber\\
 +\left(\frac{9t}{\tau}-2|a|^2-1\right)v(t)=0,
\end{align} 
with $v(t)=u^{(d)}(t/\tau)$.
As $\tau\to\infty, |a|\to 0$ under the condition $|a|^2\tau\to 1$, we obtain a confluent HE:
\begin{equation}
 t(t-1)\frac{d^2 v}{dt^2}+\left(2t-\frac{1}{2}\right)\frac{dv}{dt}+\frac{1}{4}v=0.\label{eq:confluent_hde}
\end{equation} 
Equation~(\ref{eq:confluent_hde}) is equivalent to the GE which was obtained from the continuous-time QW (see Eq.~(\ref{eq:ctqw_gde})).
This result corresponds to the result in Strauch~\cite{strauch2006connecting}.
We can confirm his result in a relation between the HE and the GE.

\section{Summary}
\label{summary}
In this section, we discuss and conclude our results.
In the present paper, we found that for the discrete-time QW, the limit density function $f^{(d)}(x)$ satisfies a HE, while for the continuous-time QW, the limit density function $f^{(c)}(x)$ satisfies a GE.
Moreover, considering the confluent HE, we confirmed a relation between the discrete- and the continuous-time QWs corresponding to the result by Strauch~\cite{strauch2006connecting}.

As a significant remark, we would like to mention a relationship between the discrete-time QW and a Schr\"odinger equation through the HE~(see \ref{app1} for detail).
The Hamiltonian of the continuous-time QW is given by the adjacency matrix of the graph on which the walk is defined~\cite{farhi1998quantum}.
However the Hamiltonian of the discrete-time QW is not known.
On the other hand, it was shown by Takemura ~\cite{takemura2003heun2,takemura2003heun,takemura2004heun,takemura2004heun2,takemura2005heun,takemura2006heun} that the HE can be transformed to the $BC_1$ Inozemtsev system which is a one-particle integrable quantum system.
In addition, the $BC_1$ Inozemtsev system includes the Calogero-Moser-Sutherland system or the Olshanetsky-Perelonov system~\cite{olshanetsky1983quantum}.
And the Hamiltonian is expressed with the Weierstrass $\wp$-function.
Thus, to find a relation among the discrete-time QW, the HE and the $BC_1$ Inozemtsev system might be one of the interesting problems.

\appendix
\section{A relation between the discrete-time QW and the $BC_1$ Inozemtsev model}
\label{app1}
In this Appendix, we transform the HE obtained in Sect.~\ref{dt} to the $BC_1$ Inozemtsev system.
The $BC_1$ Inozemtsev model is known as an integrable one-particle quantum system including the Calogero-Moser-Sutherland system (see \cite{takemura2003heun2,takemura2003heun,takemura2004heun,takemura2004heun2,takemura2005heun,takemura2006heun}).
The Hamiltonian $H$ of the $BC_1$ Inozemtsev model is defined by
\begin{equation}
 H=-\frac{d^2}{dx^2}+\sum_{i=0}^{l_j}(l_j+1)\wp(x+w_j),
\end{equation}
where the function $\wp(x)$ is the Weierstrass $\wp$-function with periods $(2w_1,2w_3)$ and the parameters $l_j\in\mathbb{C}\,(j=0,1,2,3)$ are constants.
We should note that $w_1,w_3\in\mathbb{C}$ are linearly independent on $\mathbb{R}$, where $\mathbb{R}$ is the set of real numbers.
By using the following transformation for Eq.~(\ref{eq:dtqw_hde}) (see Takemura~\cite{takemura2004heun,takemura2004heun2} for detail),
\begin{equation}
 z=\frac{\wp(x)-\wp(w_1)}{\wp(w_2)-\wp(w_1)}
\end{equation}
and putting $g(x)=u^{(d)}(z)z^{1/4}(z-|a|^2)^{3/4}$, we can get
\begin{equation}
 Hg(x)=\frac{2-|a|^2}{12}\left\{\wp(w_1)-\wp(-w_1-w_3)\right\}g(x),
\end{equation}
where
\begin{equation}
 H=-\frac{d^2}{dx^2}-\frac{1}{4}\wp(x)+\frac{3}{4}\wp(x-w_1-w_3)+2\wp(x+w_3).\label{eq:BC1_hamiltonian}
\end{equation}
The operator $H$ is the Hamiltonian of the $BC_1$ Inozemtsev model with $l_0=-\frac{1}{2},\, l_1=0,\, l_2=-\frac{3}{2},\, l_3=1,\, w_0=0,\, w_2=-w_1-w_3$, where $w_1,w_3$ are arbitrary non-zero complex numbers and linearly independent on $\mathbb{R}$.
Therefore the eigenvalue of the Hamiltonian $H$ determined by Eq.~(\ref{eq:BC1_hamiltonian}) is $\frac{2-|a|^2}{12}\left\{\wp(w_1)-\wp(-w_1-w_3)\right\}$ and the eigenfunction corresponding to the eigenvalue is the above mentioned function $g(x)$.
Via the HE, we found that the limit density function $f^{(d)}(x)$ of the discrete-time QWs is related to the $BC_1$ Inozemtsev model.

\section{Relations among measure of the QW, the HE and the GE}
\label{app2}
In this Appendix, we focus on both limit measures $f^{(d)}(x)\,dx$ and $f^{(c)}(x)\,dx$ and discuss a relation among the QW, the HE and the GE.
Putting $x^2=z$ for the measure $f^{(d)}(x)\,dx$ of the discrete-time QW, we get
\begin{equation}
 f^{(d)}(x)\,dx=\left\{\begin{array}{ll}
		 \frac{\sqrt{1-|a|^2}}{2\pi(1-z)\sqrt{|a|^2z-z^2}}\,dz & (x>0), \\[2mm]
			-\frac{\sqrt{1-|a|^2}}{2\pi(1-z)\sqrt{|a|^2z-z^2}}\,dz & (x<0).
		       \end{array}\right.
\end{equation}
The function $w^{(d)}(z)=\sqrt{1-|a|^2}/2\pi(1-z)\sqrt{|a|^2z-z^2}$ satisfies the HE with $\alpha=\beta=2,\gamma=\frac{3}{2},\delta=2,\epsilon=\frac{3}{2},q=\frac{3|a|^2+2}{2},\theta=|a|^2$:
\begin{equation}
 \frac{d^2}{dz^2}w^{(d)}(z)+\left(\frac{\frac{3}{2}}{z}+\frac{2}{z-1}+\frac{\frac{3}{2}}{z-|a|^2}\right)\frac{d}{dz}w^{(d)}(z)+\frac{4z-\frac{3|a|^2+2}{2}}{z(z-1)(z-|a|^2)}w^{(d)}(z)=0.\label{eq:dtqw_hde2}
\end{equation}
In the case of the continuous-time QW, putting $x^2/4\nu^2=z$ for $f^{(c)}(x)\,dx$, we obtain
\begin{equation}
 f^{(c)}(x)\,dx=\left\{\begin{array}{ll}
		\frac{\nu}{\pi\sqrt{z-z^2}}\,dz &(x>0),\\
		 -\frac{\nu}{\pi\sqrt{z-z^2}}\,dz &(x<0).
		       \end{array}\right.
\end{equation}
For the function $w^{(c)}(z)=\nu/\pi\sqrt{z-z^2}$, the following GE with $\alpha=\beta=1,\gamma=\frac{3}{2}$ is realized:
\begin{equation}
 z(z-1)\frac{d^2}{dz^2}w^{(c)}(z)+\left(3z-\frac{3}{2}\right)\frac{d}{dz}w^{(c)}(z)+w^{(c)}(z)=0.\label{eq:ctqw_gde2}
\end{equation}
Note that we can derive Eq.~(\ref{eq:ctqw_gde2}) from Eq.~(\ref{eq:dtqw_hde2}) in a similar fashion as in Sect.~\ref{dtct}.
Moreover Eq.~(\ref{eq:dtqw_hde2}) can be transformed into the eigen equation of the $BC_1$ Inozemtsev model with $l_0=-\frac{1}{2}, l_1=-1, l_2=-\frac{3}{2}, l_3=-1, w_0=0, w_2=-w_1-w_3$ and the eigenvalue $\frac{2-|a|^2}{12}\left\{\wp(w_1)-\wp(-w_1-w_3)\right\}$.


\vspace{5mm}

\noindent Norio Konno: Department of Applied Mathematics, Faculty of Engineering, Yokohama National University, Hodogaya, Yokohama 240-8501, Japan,
\email{konno@ynu.ac.jp}\\

\noindent Takuya Machida: Research Fellow of Japan Society for the Promotion of Science,
\email{machida@stat.t.u-tokyo.ac.jp}\\

\noindent Tohru Wakasa: Department of Basic Sciences, Kyushu Institute of Technology, Tobata, Kitakyusyu 804-8550, Japan,
\email{wakasa@mns.kyutech.ac.jp}

\end{document}